\documentstyle{article}

\begin{document}

\begin{center}
{\bf A MORE SENSITIVE LORENTZIAN STATE SUM}

\bigskip

Louis Crane and David Yetter\\
Dept. of Mathematics\\
Kansas State University \\
Manhattan, KS 66503

{\it crane@math.ksu.edu, dyetter@math.ksu.edu}
\end{center}

\bigskip

{\bf ABSTRACT} {\bf \it We give the construction modulo normalization of a new state sum model for
lorentzian quantum general relativity, using the construction of
Dirac's expansors to include quantum operators corresponding to edge
lengths as well as the quantum bivectors of the Barrett Crane model,
and discuss the problem of its normalization. The new model gives rise
to a new picture of quantum geometry in which lengths come in a
discrete spectrum, while areas have a continuum of values.}

\bigskip

{\bf I. INTRODUCTION}

\bigskip

The state sum models for quantum general relativity in [1,2]
have attracted considerable attention, especially since the discovery
of their finiteness in [3]. Unfortunately, they seem to have two
weaknesses. The first is the well noted difficulty of finding a
classical limit for them, the second is the fact that they contain an
excessively large contribution from degenerate geometries [4,5,6].

In hindsight, it is clear why such problems should
arise. The models do not have quantum operators corresponding to edge
lengths or components of edge vectors. In the classical derivation of the model, the
constraints on the bivectors implied that edge lengths would exist,
except in degenerate configurations. This led some researchers to hope
that it would be possible to have ``edges without edges,'' i.e. to
recover the geometry of the edges as a side effect of the
constraints. The discovery of the degenerate configurations seems to
indicate that this hope was misplaced.

The problem now arises of how to modify the model in such a way as to
retain its positive features, while removing the degenerate
configurations, and being able to impose more stringent matching
conditions between adjacent 4-simplices, thus providing a more natural
classical limit.

One promising  approach to this problem was proposed in [6].
This approach may be described as conservative, in the sense that it
retains the basic form of the model, as a multiple integral in
hyperbolic space, while projecting out the dangerous region of the
integral.

We want to propose a ``radical'' approach. We construct a state sum
model which has both edge length and bivector operators. The new model
will use the expansors of Dirac [7], as quantizations of the edges of
a triangulation, decompose their tensor products to obtain unitary Lorentz
representations to represent quantizations of the bivectors on the
faces as in the model of [2], then combine them in 10J symbols as in
the BC model. The objective is to retain the dynamics of the BC model,
while enlarging the variables to allow more precise joining conditions
between simplices.

The expansors of Dirac can be thought of as a relativistic analog of
the solutions of the spherical harmonic oscillator. Much as the
eigenvectors of the spherical harmonic oscillator form a basis for
$L^2(R^3)$, the expansors give a basis for $L^2(R^4)$, and can be
considered a quantization of an edge displacement in a four
dimensional simplex.

Very roughly, the physical
idea is to augment the model by including solutions of the Schrodinger
equation with radial terms
analogous to the solutions for the hydrogen 
atom instead of just the angular pieces which correspond to spins 
(the harmonic oscillator in 3D as a radial
potential, is very
similar to, though not identical with, the hydrogen atom since they
are both radial potentials). This is
consonant with the intuition that the triangulations of this model are
``atoms of geometry''. Motivated by this intuition, we will explore
using the oscillator hamiltonian to normalize the theory.

Another intuitive way of understanding the new model is that the
models of [1,2], were quantizations of a description of the
geometry of a triangulated manifold by the bivectors on its
2-simplices, which can be related to the infinitesimal generators of
the rotation group. The new model is a quantization of a description
of the displacement vectors on edges and bivectors on 2-simplices, which
can  be related to the full set of generators of the poincare group,
translations on edges, rotations on faces.

Our proposal can also be understood in a categorical context.
The various state sum models for quantum GR are all constrained
versions of categorical state sums related to topology [8].
From the higher categorical standpoint [9,10], it is rather odd that
we are using only a tensor category, rather than a tensor 2-category
in a 4D state sum. Indeed, [11,12] has shown that a state sum in
4D similar to the ones in [8], can be constructed from a  ``spherical'' 2-category. The
most obvious difference between the two types of state sums is that in the
2-categorical approach both edges and faces get information.

The new state sum we are proposing is related to a new
2-category [13] which corresponds to the combination of the poincare and
lorentz groups into a categorical group [14,15]. The construction
makes use of a new class of categorical structures called ``measured
categories.'' Indeed, it is the coincidence of a natural higher
categorical picture with some natural quantum geometry which
motivates the model we will be describing. The categorical aspects
of this construction will appear in a companion paper [13], the present
paper, intended for an audience in mathematical physics, will not
discuss them. 

Since the expansors are no longer familiar, we will begin with an
exposition of them, which should be self contained for anybody
familiar with the unitary lorentz representations which play a role in
[2]. Then we will outline the construction of the new model.

\bigskip

{\bf II. THE EXPANSORS }

\bigskip

In 1945, Dirac [7] introduced the first study of unitary representations
of noncompact lie groups by constructing a family of examples, which
he called the expansors, for the lorentz group.

His construction is a lorentz signature analog of the three
dimensional spherical
harmonic oscillator.  Let us first review the construction in the
three dimensional euclidean signature case. We can take as
representing spaces the homogeneous polynomials of degree n in the
formal variables $ \xi_x, \xi_y, \xi_z $.

The action on this space of the three dimensional rotation group SO(3) is
given by substitution of the rows of the matrices for group elements
into the polynomials. A simple calculation shows that such a
substitution preserves the inner product on the space of polynomials
in which monomials with different exponents are orthogonal and each
monomial ${\xi_x}^n {\xi_y}^l {\xi_z}^m $ has length $( n!l!m!)^{1/2}$. Thus the
expansors of each degree form a unitary representation of SO(3).

 These representations are not irreducible.  In general, the
 homogeneous polynomials of degree n  decompose into a direct sum of
 one copy each of every irreducible representation whose parity is the
 same as n from 0 or 1 up to n. This fact is well known in nuclear
 physics. This result is closely analogous to the well known solution
 of the hydrogen atom, whose three quantum numbers would correspond to
 the degree n, the spin of the representation, and the z component of
 the spin.

The relationship between the representations so constructed, which
Dirac called expansors because they come from an expansion of a
polynomial under a substitution, and the solutions of the spherical
harmonic oscillator, is that the $ \xi$ variables correspond to the
raising operators for the three harmonic oscillators in the three
dimensions.

\bigskip

$ \xi_x = x+d/dx$ etc.

\bigskip

The inner product given on the basis above then corresponds to the
ordinary inner product in $L^2(R^3)$ as a simple calculation
shows. The degree of the polynomial is just the energy level of the
solution.

This construction is not original to Dirac. His contribution was
extending the construction to the lorentzian signature, while
preserving the unitary nature of the representations.

Before we discuss the lorentzian signature result, we will briefly
describe the four dimensional analog of this construction, because it
may well prove interesting in the study of euclidean quantum gravity.

The construction in 4D is the same as in 3D except we must include a
fourth $\xi$ variable. In complete analogy to 3D; we thus obtain for
each n a reducible unitary representation of SO(4).

 The decomposition
into irreducibles is similar, with the interesting new feature that
only the balanced representations of SO(4) now appear.
For the case of
SO(4), this means that the two half integers indexing the
representation are equal.
This is not
surprizing, since only they appear as harmonics on
$S^3$. Nevertheless, for readers familiar with the development to date
in the state sum models, it is already clear that certain
simplifications are going to ensue.

In order to use a similar construction to obtain unitary
representations of SO(3,1), Dirac hit on the expedient of including a
fourth parameter $\xi_t$, and considering the homogeneous polynomials
in $\xi_x, \xi_y, \xi_z,$ and $(\xi_t)^{-1}$. This is no longer a finite
dimensional space for any positive or negative degree, as is
necessary, since SO(3,1) is noncompact. Assigning a length

\bigskip

$(l!m!n!)^{1/2}(p!)^{-1/2}$

\bigskip

\noindent to the monomial

\bigskip

${\xi_x}^l{\xi_y}^m{\xi_x}^n{\xi_t}^{-p}$

\bigskip

\noindent we again find that SO(3,1) preserves the inner product on the space of
homogeneous expansors of degree n, which we denote $E^n$.

This discovery of Dirac's was the first example of a family of unitary
representations of a noncompact group to appear. Just as in the
Euclidean case, the representations compose into series of
irreducibles.

The irreducibles which appear are in the principal series [16,17] of
representations $R(n, \rho)$  of SO(3,1) for n integer and $\rho$
real. Interestingly, only the balanced representations R(n,0) and
R(0,k) appear in them.  Negative degree extensors decompose as strings
of [n/2] R(i,0)'s of the same parity as n, followed by all the
R(0,i)'s, while the positive degree extensors decompose into strings
of R(0,i)'s only, beginning with i=n and containing all the R(0,i)'s
of appropriate parity. These facts are easy to check, comparing the
formulas in [7] and [16].

The expansors have an interpretation in terms of the tensor product of
4 harmonic oscillators with the oscillator in the t direction assigned
a negative energy. The degree is then the energy level of the
solution.

Using the eigenbasis for the ``lorentzian harmonic oscillator'', we can
thus quantize minkowski space in the form

\bigskip

$L^2(M^4) = \oplus E^n$;

\bigskip

\noindent with the further decomposition as representations of SO(3,1) following immediately.

The positive and negative parts of our sum quantize the timelike and
spacelike regions. We believe this could be treated using geometric quantization.

\bigskip

{\bf III. THE QUANTUM TRIANGLE}

\bigskip

Now we have a tool for constructing a quantized description of the
geometry on a lorentzian simplex. We first need to see how to combine data on a
triangle to produce quantum edges and a quantum bivector.

The classical conditions we need to impose are that the vector sum of
the displacements on the three edges is 0, and that the bivector is the
wedge product of any two of them, with a sign depending on
orientation. It is actually slightly easier to order the 3 vertices of
the triangle 1,2,3, orient the edges by ascending order, and require the
vector on the edge labelled 13 to be the sum of the other two vectors.

It is easy to see how to quantize this. The space $T_q=(\oplus
E^n)^{\otimes 2} $ is naturally identified with the space $ T_S=L^2[
\{ (a,b,c) \; | \;
a,b,c \epsilon M^4 ;\; c=b+a\}] $ . We can
restrict to the skew symmetric part of $T_q$ . Passage from one
ordering to another would give us a different description of the same
space. Explicitly,
\bigskip

$ \sigma_q ;L^2[a,b]  \leftrightarrow L^2[a,c] $ for $c=b+a$.

\bigskip

Furthermore, this map from one parametrization of $T_S$ to
another is equivariant with respect to the action of
SO(3,1), so the decomposition of the skew part of $T_q$  would still
give the same combination of irreducible representations of SO(3,1)
after a change of ordering.  Thus, the sum of representations of
SO(3,1) occurring in ${T_q}^A$ is a natural setting for a quantization of the
bivector in the quantum triangle. The action of the operators
corresponding to components of the bivector is just the action of the
lie algebra elements on the representations, as in the correspondence
in [1,2] between the lie algebra and the bivectors.

We now have a space, $T_S$ identified with its presentation $T_q$,
which is a quantization of a minkowskian triangle, while its skew
symmetric part is a natural setting for the quantum bivector.

How should we extract a ``quantum bivector?'' To see this, we should
think about the geometry of the situation where two vectors are wedged
into a bivector. The map
\bigskip

 $ \pi: v_1 \times v_2 \rightarrow v_1 \wedge v_2$

\bigskip

\noindent is a fibration with fiber $R^2 \times S^1 $, since the set of vectors
with a given wedge is the set of pairs in the plane which make a
parallelogram with a given area. The image of $\pi$ is the set of
simple bivectors of $M^4$.

We want a quantization of the image of $\pi$, so we want to identify
the representations in the decomposition of ${T_q}^A$ which pass down
to the image. The answer is interesting, if familiar. Ths $S^1$ piece
of the fiber of $\pi$ is rotated by the rotation in the plane of the
bivector to which it is sent. Thus, the part of the function space on
the fiber which is constant along $S^1$ transforms the same way as the
corresponding function on the image.

Now, which representations correspond to constants along $S^1$, and
which to higher Fourier components? The little group of a bivector is
contained in a maximal torus of SO(3,1), so the representations of the
principal series are induced from it. Tensoring with a nontrivial
fourier component will therefore change a representation $R(0,\rho)$
to $R(k, \rho)$.

\bigskip

{\bf { \it  Thus, in order to make a quantum description of the
bivector on the quantum triangle, we must project out the unbalanced
representations appearing in ${T_q}^A$.}}

\bigskip

This result is not surprizing, since we should only obtain simple
bivectors when we wedge two vectors, but it is reassuring that it
emerges naturally from our categorical quantization procedure. We will now have an
infinite overcounting of the representations which survive in the
image of $\pi$, from the tensor product with the $L^2(R^2)$ piece from
the fiber of $\pi$. It is already clear from the definition of $T_q$
that it will contain infinitely many copies of each representation.

Let us describe more explicitly the representation of the quantum
bivector in the quantum triangle. We are taking the skew part of the
tensor product of the direct sum of all the expansors $E^n$. Each of
these decomposes into a tower of copies of $R(0,{\rho}_i )$ for all
integer valued of $\rho$ beginning with n and of the right parity. We
then project onto the balanced part, ie. only the copies of
$R(0,\rho)$ , where now $\rho$ is any positive real number. Thus we
get copies of the direct integral of all the $R (0,\rho)$ for each
combination of two indices n, and two ${\rho}_i$, skew symmetrized
with respect to the pair of indices. The n indices are the quantum
version of the length operators on the sides. The other index
${\rho}_i$ is analogous to the angular momentum quantum number for the
hydrogen atom. We do not yet understand its quantum geometry.

In this construction, we are using two different operations on the
expansor space. One is the identification $\sigma_q$ of  $L^2[a,b]$  with $L^2 [a,c]$
 above, the other is the skew symmetric tensor product of
representations of SO(3,1). These are quantum versions of vector
addition and the wedge product of vectors. The  fact that our expansor space can both be naturally added and
wedged, which we are exploiting here, is related to the categorical
idea mentioned above, that representations of the poincare and lorentz
groups should be fused into a higher categorical structure. $T_q$ and its
transformation $\sigma_q$ are the same space and transformation which occur
in the 2-categorical picture, where the natural decomposition
of $L^2(M^4)$ is as a direct integral of characters of the
translation group, and $\sigma_q$ is simply tensor product of
characters [13, 14, 15].  This view makes even clearer the fact that
$T_q$ is a good quantization of the triangle:  tensor product of characters
of the translation group is simply vector addition.

\bigskip

{\bf IV. AN OUTLINE OF THE NEW MODEL}

\bigskip

Now that we have our basic toolkit, it is easy to see how to set about
constructing a new model. We attach one copy of $T_q$ to each triangle
in the triangulation, decompose the alternating part of each into
irreducible representations of SO(3,1) by tensoring the decompositions
of each $E^n$, project out all except the balanced representations,
form regularized 10J symbols from them as in [2], and form the sum
over all choices of irreps in the triangulation of the product of all
10J symbols. We restrict our sum to terms in which all edges in four
simplices which are incident in the triangulation have the same
$E^n$. This is a quantization of requiring all matching edges to have
the same lengths, all triangles to close, and the bivectors on faces
to correspond to the edges.

At the price of vastly increasing our degrees of freedom, we now have
a state sum with quantum edge variables which match between adjacent
4-simplices, and a dynamics which will reproduce the Einstein-Hilbert
action by the usual argument.

We conjecture that the degenerate states of the old model [6] will now
be of measure 0, since they do not correspond to any assignment of
edge lengths. This is a rather subtle conjecture, since some global
constraints must appear among the quantum bivector data on a 4-simplex.

Of course, in this naive unnormalized version of the model, the sum
will be infinite. We discuss issues of normalization below.

It is interesting that the new model is a more complicated sum of
combinations of balanced representations only, with more projections of
tensor products of balanced representations onto balanced
representations. This means it is built out of the same mathematical
pieces as the B-C model [2], and can therefore be represented as
multiple integrals of Feynman type over $H^3$ [18, 19].
This is an optimistic sign that it will have a good normalization.

To see more explicitly what the new model will look like, it is a good
idea to pick an ordering of the vertices of a four simplex. We can
then choose to represent the copies of $T_S $ on the six triangles
incident on the vertex labelled 1 by the copy of $T_q$ corresponding
to the two edges incident on 1, in lexicographic order. This means starting with 4
copies of $\oplus E^n$, one on each of the four edges out of vertex 1,
decomposing into representations of SO(3,1), and forming alternating
tensor products in pairs. This can be thought of as a ``quantum
frame''. The other four triangles can
be represented in terms of the representations  on the four edges by
using the transformation $\sigma_q$  of $T_S$ corresponding to a reordering. In
further developing the model it will be necessary to give an explicit
form for $\sigma_q$ . It is analogous to the problem of writing a plane wave
in terms of atomic orbitals, so familiar from scattering theory.

This procedure gives a map

\bigskip

$ \mu : { (\oplus E^n)}^{\otimes 4} \rightarrow \oplus R(0,\rho
)^{\otimes 10} $

\bigskip

which will need to be studied carefully to test the above conjecture.

\bigskip

{\bf V. ISSUES OF NORMALIZATION; SMALL ATOMS }

\bigskip

Our first suggestion in regard to normalization relates to a change in
point of view as to the meaning of terms in our quantum state sum.

We want to think of the quantum data on our 4-simplices as
``atoms of geometry,'' [20] rather than as possibly large regions which
happen to be utterly flat. We have several reasons for this. In the
first place, the idea of spacetime as a nondenumerably infinite point
set seems inextricably wedded to classical physics. We believe the
point of our spin foam models is to replace this picture with a
superposition of discrete structures. In the second place, a large
volume which is treated as completely flat is unphysical. We would
also mention the unpleasant behavior of spin foam models in the large
J limit as something to be avoided.

Motivated by the above considerations and by the connection between
the decomposition of $L^2(R^4)$ we are using and the spherical
harmonic oscillator hamiltonian, we propose to put another factor in
our state sum. This would be a product of $e^{-H}$ for each edge in
the triangulation. Actually 2 choices are possible, we can either make
the edges spacelike, or make them timelike and use the opposite sign
convention in our definition of $H= x^2+y^2+z^2-t^2$. In the expansor
picture, this amounts to an exponentially damped weight on $E^n$
terms.

Certainly, this would help a lot with the divergence. More delicate
computation will be needed to see if it suffices.

 It seems clear that
the problem of the multiplicity of copies of $R(0,\rho)$ due to the
indices n would be controlled by this ansatz. The other multiplicity
needs to be better understood. It appears to be an artifact caused by
the overcounting from the extra degrees of freedom in the fibration
discussed above, but a nice way to cancel it is needed.

We ought to remind ourselves that the initial normalization of the BC
model, motivated by the connection with TQFT, turned out to be not the
best one, and that it was a rederivation of the model as an expansion
in feynman diagrams that gave the correct one [21].

This poses the question of finding some feynmanological picture for
the new model. We believe this is an interesting question for several
reasons.

In the first place, the integrals in the BC model turned out to be
essentially feynman integrals in curved space. The new model will be
composed of slightly more complicated integrals of a similar type.

Secondly, feynman diagrams are essentially morphisms in a tensor
category. The new model is 2-categorical in a geometrically natural
way, so it should be a good place to try to understand the
2-categorical analog of feynmanology, which should be important in
advancing the spin foam picture.

\bigskip

{\bf VI CONCLUSIONS}

\bigskip

Let us restate the construction of the model as far as we currently
understand it. We begin by summing over an assignment of one $E^n$  to
each edge in the triangulation. We then put a normalization factor of
$e^{-n}$ on each edge, impose the projection onto $T^q$ for each
triangle of each 4-simplex, decompose the skew tensor product on each
triangle into irreps of the lorentz algebra, project onto the balanced
irreps, form the B-C 10J symbols,
and take the integral of products of terms. It is not clear yet how much
our constraints help us with our naive infinities, or whether further
normalization is necessary. We believe a deeper study of the new
quantum geometry of this model will clarify this.

The fibration picture for the quantum triangle suggests that the naive
divergences in this model just come from an infinite multiplicity
corresponding to the volume of $R^2 \times S^1$. This makes us
optimistic that a consistent regularization can be found, much as the
divergences in the model of [2] were all controlled by dividing out a
single infinite volume.

The hope is that the classical states (terms of stationary phase) in
this sum will be discrete einstein metrics with all simplices very
small.

We note that the idea of [22], that we need to divide by an infinite
diffeomorphism volume factor to regularize a state sum, is formally
similar to the current situation. We are not sure if there is a deeper connection.

Since there is some interest in the question of discrete versus
continuous spectra for geometrical quantities in quantum general
relativity, let us note that the length variables in our model are
discrete because of the decomposition of the expansor spaces described
above, but that the area spectrum will be continuous because the
decomposition of the tensor product of the irreducible unitary representations
takes the form of a direct integral, not a discrete sum.

It is also interesting to consider the possibility of a q-version of
the present model. Since it is well known that a natural q-harmonic
oscillator exists [23], it is easy to see how to go about constructing q-expansors.
A q-version of the B-C model has been studied in [24]. It would be
interesting to study the decomposition of the q-expansors, and to see
if they led to a manifestly finite version of the 2-categorical model.

At  this point, it is still rather speculative to be thinking about
adding matter to this system. Nevertheless, we point out that we are
only including a factor which exponentially damps the length on each
edge in our preliminary proposal for a normalization. This is in
contrast to state sum models for TQFTs in which rather subtle terms
arise, which would not make sense for nonmanifold configurations. It
is therefore possible to consider the suggestion of conical matter [25]
in the model of this paper.

\bigskip

{\bf ACKNOWLEDGEMENTS} {\it This paper was strongly motivated by
  conversations with John Barrett and Marco Mackaay, who were kind
  enough to show one of us reference 14 in an unfinished state.}

\bigskip

{\bf BIBLIOGRAPHY }

\bigskip
1. J. W. Barrett and L. Crane,  Relativistic spin nets and quantum
   gravity, J. Math. Phys. 39 No. 6 (1998) 3296-3302

\bigskip

2. J. W. Barrett and L. Crane, A lorentzian signature model for quantum
   general relativity Class. Quant. Grav. 17 No 16 (2000) 3101-3118

\bigskip

3.L. Crane A. Perez and C. Rovelli, A finiteness proof for the
  lorentzian state sum model for quantum genreral relativity. gr-qc 0104057

\bigskip

4.J. C. Baez, . D. Christensen, and G. Egan, Asymptotics of 10J
  symbols, Class. Quant. Grav. 19 (2002) 6489

\bigskip

5. J. W. Barrett and C. M. Steele, Asymptotics of relativistic spin
   networks, gr-qc 0209023

\bigskip

6.L. Freidel and D. Louapre, Asymptotics of 6J and 10J symbols, hepth 0209134

\bigskip

7.P. A. M. Dirac, Unitary representations of the lorentz group,
  Proc. Roy. Soc. A. 183 (1945) 284-295.

\bigskip

8. L. Crane, L. H. Kauffman and D. Yetter, State sum invariants of
  4-manifolds, JKTR 6(2) 1997 177-234

\bigskip

9.J. Baez, Higher dimensional algebra and plank scale physics, gr-qc
  9902017

\bigskip

10.L. Crane and I. B. Frenkel, Four dimensional topological quantum
   field theory, Hopf categories, and the canonical bases, JMP 35
   No. 10, (1994) 5136-5154

\bigskip

11. M. Mackaay Spherical 2-categories and 4-manifold invariants
    Adv. Math. 143 (1999) 288-348

\bigskip

12. M. Mackaay Finite groups, spherical 2-categories and 4-manifold
    invariants Adv. Math. 153(2) (2000) 353-390

\bigskip

13L. Crane and D. Yetter,  Measured Categories and Representations of
2-Groups, in preparation

\bigskip

14.J. W. Barrett and M. Mackaay, Categorical representations of
   categorical groups, to appear.

\bigskip

15. J. Baez, Higher Yang Mills theory, hep-th 0206130

\bigskip

16. Harish-Chandra, Infinite irreducible representations of the
    lorentz group, P. Roy Soc. A. Math, 189 (1947) 372

\bigskip

17. I. M. Gelfand, I. M. Graev, and N. Vielkin, Generalized Functions,
    vol 5, Academic Press 1966

\bigskip

18. J. Barrett, The classical evaluation of relativistic spin nets,
    Adv. Th. Math. Phys. 2 (1998) 593-600

\bigskip

19. L.  Freidel and K. Krasnov, Simple spin networks as feynman graphs,
    J. Math. Phys. 41 (2000) 1681-1690.

\bigskip

20 M. Reisenberger, private conversation.

\bigskip

21. R. De Pietri, L. Freidel, K. Krasnov and C. Rovelli, Barrett Crane
    model from a Boulatov-OOguri field theory over a homogeneous space
    Nucl. Phys. B574 (2000) 785-801. 

\bigskip

\bigskip

22. L. Freidel and D. Louapre, Diffeomorphisms and spin foam models
    gr-qc 0212001

\bigskip

23. A. J. Macfarlane, Qn q-analogues of the harmonic oscillator and
    quantum group $SU(2)_q$ J. Phys. A 22 (1989) 4581

\bigskip

24.K. Noui and P. Roche Cosmological deformations of lorentzian spin
   foam models, gr-qc 0211109

\bigskip

25 L. Crane, A new approach to the geometrization of matter, gr-qc 0110060

\end{document}